\documentclass[intlimits,twoside,a4paper]{article}

\usepackage[cp1251]{inputenc}

\usepackage{cmpj3}

\def\a{\alpha}
\def\b{\beta}
\def\g{\gamma}
\def\e{\varepsilon}
\def\d{\delta}
\def\k{\kappa}
\def\l{\lambda}
\def\m{\mu}

\def\t{\tau}

\def\o{\omega}
\def\x{\xi}
\def\h{\eta}
\def\r{\rho}
\def\s{\sigma}

\def\S{\Sigma}
\def\G{\Gamma}
\def\D{\Delta}
\def\O{\Omega}

\def\ua{\uparrow}
\def\da{\downarrow}

\def\bk{\mathbf{k}}

\def\bq{\mathbf{q}}
\def\br{\mathbf{r}}
\def\bp{\mathbf{p}}
\def\bn{\mathbf{n}}

\def\bs{\mathbf{s}}
\def\lGF{{\langle\langle}}
\def\rGF{{\rangle\rangle}}
\def\be{\begin{equation}}
\def\ee{\end{equation}}
\def\bea{\begin{eqnarray}}
\def\eea{\end{eqnarray}}
\def\nn{\nonumber}
\def\lb{\label}

\DeclareMathOperator{\Tr}{Tr}
\DeclareMathOperator{\arctanh}{arctanh}
\DeclareMathOperator{\arcsinh}{arcsinh}
\issue{2018}{21}{3}{33704}
\doinumber{10.5488/CMP.21.33704}

\begin{document}

\title{On the theory of high-$T_\text{c}$ superconductivity of doped cuprates}
\author{Y.G. Pogorelov\refaddr{Porto}, V.M. Loktev\refaddr{BITP,KPI}}
\addresses{
\addr{Porto} IFIMUP-IN, Departamento de F\'{i}sica e Astronomia, Universidade do Porto, Porto, Portugal
\addr{BITP} Bogolyubov Institute for Theoretical Physics of the National Academy of Sciences of Ukraine, \\ 14b Metrologichna St., 03143 Kyiv, Ukraine
\addr{KPI} National Technical University of Ukraine ``Igor Sikorsky Kyiv Polytechnic Institute'', \\ 37 Peremohy Ave., 03056 Kyiv, Ukraine
}

\date{Received June 12, 2018, in final form July 23, 2018}

\maketitle

\begin{abstract}
A theoretical analysis is presented on possible effects of disorder by dopants in high-temperature superconducting cuprate perovskites, to define their basic spectra of spin and electronic excitations, and the subsequent observable properties, especially doping dependence of superconducting order parameter. The central point in the proposed physical picture is formation of specific impurity subband within the insulating bandgap of initial undoped material, serving as a source for the system metallization and further transition into superconducting state with anisotropic order parameter.
\pacs 71.30.+h, 71.35.-i, 74.0.-z, 74.20.Rp, 74.62.Dh, 74.72.-h
\end{abstract}

\section{Introduction}
The phenomenon of high-$T_\text{c}$ superconductivity discovered more than 30 years ago in layered perovskite cuprates \cite{bed,wolf} 
and then found in many other materials \cite{kam,wat,droz} still presents a challenge for the condensed matter theory. Besides 
its main distinctive feature of critical temperature $T_\text{c}$ for superconducting (SC) order to exceed the theoretical limits established earlier 
for ``classical'' superconductors \cite{ginz}, it reveals a lot of other differences from the canonical Bardeen-Cooper-Schrieffer (BCS) picture 
\cite{bcs} such as: an extremely short SC coherence length $\xi$ and respective huge values of the Ginzburg-Landau parameter $\k$, 
unusual $d$-wave symmetry of the SC order parameter, onset of the SC order under doping of initially insulating materials (as 
La$_2$CuO$_4 \to$ La$_{2-c}$Sr$_c$CuO$_4$) and then a non-monotonous $T_\text{c}$ variation with growing doping rate $c$, rising from 
zero at the metallization threshold $c_{\text{met}}$ to reach a maximum and then decaying to vanish at some upper limit $c_{\text{cr}}$ \cite{kei}. These 
exotic properties were deeply studied in various untraditional theoretical ways, with the main accent on strong electronic correlations at the 
transition from insulating to metallic state, beginning from the Anderson resonating valence bond model \cite{and,and2,bask}, 
Zhang-Rice singlet model \cite{zhr}, Emery-Kivelson phase fluctuation model \cite{emk} and many others. For instance, a recent study by this 
approach suggested an enhanced stability of the SC order in high-$T_\text{c}$ cuprates under the doping induced disorder \cite{tang}. Nevertheless, 
some issues in this field still remain out of scope of the existing approaches, for instance recent experimental findings of anomalously long 
magnetic penetration depth $\l$ and hence anomalously low superfluid density $n_s$ in the overdoped ($T_\text{c}$ decaying) SC range \cite{boz}, 
in a striking contrast with the BCS theory predictions \cite{boz1}. 

It should be noted here that all the above mentioned theoretical pictures of SC state, either canonical or strong-correlational, consider many-body 
processes in high-$T_\text{c}$ materials, even in presence of randomly located dopants, as in uniform crystalline systems with full translational invariance, 
implicitly disregarding the effects of static disorder. On the other hand, such effects can qualitatively modify spectral and related observable properties 
of various crystalline systems with disorder as was shown by many researchers, beginning from the seminal studies by Lifshitz on emergence of local 
modes and impurity bands \cite{lif,lgp} and by Mott on mobility thresholds between localized and band-like states \cite{mott,mott1}. In 
particular, restructuring of excitation spectra in high-$T_\text{c}$ materials under perturbations from both dopants and isovalent impurities was discussed in 
detail by the authors \cite{lp,lp1}, though leaving aside the role of these perturbations for the very formation of the SC order. The present work is 
just aimed at extension of our earlier theoretical approaches to this latter aspect of the high-$T_\text{c}$ problem, in order to clarify the decisive importance 
of disorder effects for understanding the physical difference between BCS and high-$T_\text{c}$ types of superconductivity. This task is being done here 
with parallel analysis of two different kinds of elementary excitations in the perovskite compound: fermion quasiparticles by electron hopping between 
neighbor Cu sites and boson-like spin excitations at strong enough indirect antiferromagnetic (AFM) superexchange between these sites via intermediate 
O sites. Their interplay is found to define the insulating state of the undoped material and then, introducing perturbations from dopant atoms on both 
excitation modes, the possibility for its transition to a specific metallic state with strong disorder effects and anomalously narrow conduction band is 
established. The latter specifics is further analyzed for its influence, together with the fermion-boson interplay, on the formation and properties of the 
SC state, in particular, the non-monotonous dependence of the SC order parameter on the doping level. 

The following consideration starts in section~\ref{gf} from definition of the basic Green functions, specific for the analysis of different types of elementary 
excitations in a disordered crystalline system. Section~\ref{mod} explains the choice of effective Hamiltonians for spin and charge degrees of freedom, 
including their perturbations due to dopants. Then, we find the excitation spectra in this system, with a special emphasis on the effects of spin dynamics 
on charge dynamics, either in the initial state or under finite doping. In the latter case, the important doping effects for decay of long range spin order 
and for metallization of charge carrier spectrum are analyzed in section~\ref{dop}. The resulting excitation spectra also contain, besides the band-like modes 
described by their dispersion laws, possible localized modes near the dopant atoms, all resumed in the corresponding density of states (DOS) function. 
In particular, this function permits to define the position of Fermi level under doping, one of central issues for the onset of SC order. Then, the characteristics 
of the resulting SC state as functions of doping and its related disorder are obtained in section~\ref{sc}, permitting to better understand the experimentally 
measured anomalies in this state, compared to the conventional BCS theory predictions. Finally, some discussion and conclusions on the developed 
SC model are presented in section~\ref{disc}.

This work is dedicated to the 80th birthday of Prof. I.V. Stasyuk, a renowned Ukrainian theorist, recognizing his very valuable contribution to the 
theory of narrow-band metals and pseudo-spin operators in the field of phase transitions in solids \cite{stas}, especially of superconducting transitions 
\cite{plak}. 

\section{Green functions and observable values}
\lb{gf}
The spin and charge dynamics in this system are analyzed below through the standard techniques of two-time Green functions (GF's) \cite{zub,bbt}:
\be
\lGF A|B \rGF_{\pm,\e} = \frac \ri\piup \int_0^\infty \re^{\ri(\e + \ri0)t}\langle [A(t),B(0)]_\pm\rangle \rd t,
\lb{GF}
\ee
where $\langle \ldots\rangle = (\Tr  \re^{-\b H})^{-1}(\Tr  \re^{-\b H} \ldots)$ is the quantum statistical average of any 
operator in the system with full Hamiltonian $H$ at inverse temperature $\b = 1/(k_{\text{B}}T)$, the commutator $[A,B]_-$ or anticommutator $[A,B]_+$ is 
chosen respectively for the operators $A$, $B$ of Bose or Fermi type (implying respective types of thermal averaging for them), while positive infinitesimal 
imaginary part $\text{i}0$ in the energy argument of Fourier transform corresponds to the retarded GF type. These functions obey the equation of motion:
\be
\e\lGF A|B\rGF_{\pm,\e} = \langle[A,B]_\pm\rangle + \lGF[A,H]_-|B\rGF_{\pm,\e}\,,
\lb{eqm}
\ee 
permitting their consequent calculation with given $H$ (in what follows, the explicit symmetry and energy indices of the GFs are typically omitted). The 
basic GF advantage is to provide directly the system observable values from the spectral theorem:
\be
\langle A B \rangle = \frac 1\piup \int_0^\infty \Im \lGF B|A\rGF_\e \rd\e.
\lb{spt}
\ee 

It should be noted that the two-time GF's defined by equations~\eqref{GF}--\eqref{spt}, though being most suitable to analyze the effects by doped charges and 
associated disorder from dopant atoms (of central importance in the present work), but are only justified in the zero temperature limit and they are mostly 
employed in this limit below. Otherwise, the case of finite temperatures should be treated, strictly speaking, with more specialized Matsubara GF's \cite{mats,eco}, though producing some technical problems for their use in disordered systems. Nevertheless,  some qualitative estimates of critical temperatures 
for both AFM and SC ordering under doping are possible even from the simpler two-time GF formalism, so we choose to stay with it for all the following study. 
To distinguish between the two different types of excitation spectra in the present system,  we shall denote the GF energy argument $\o$ for Bose-type spin 
states and $\e$ (referred to the chemical potential $\m$) for Fermi-type charge carrier states.

\section{Model Hamiltonians and excitation spectra}
\lb{mod}
\begin{figure}[!b]
	\center \includegraphics[width=8cm]{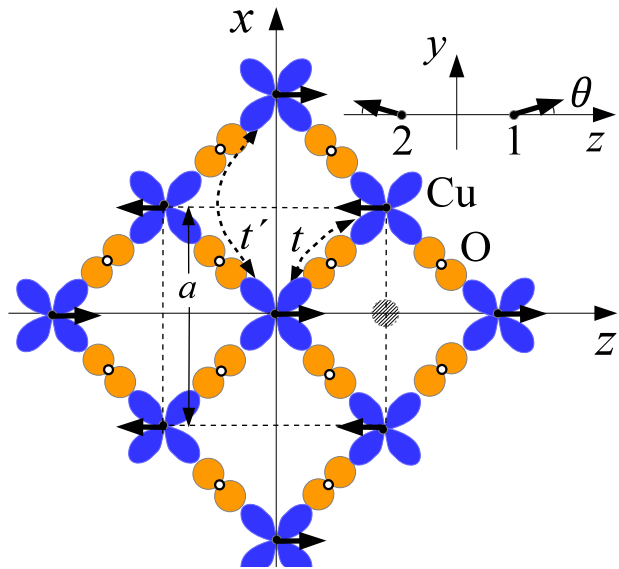}\\
	\caption{(Colour online) An elementary cell (dashed lines) in the CuO$_2$ plane with AFM ordered Cu$^{2+}$ spins (shown by arrows). Indirect hoppings (permitted at 
		presence of doped holes), $t$ between nearest neighbor Cu $d$-orbitals and $t'$ between next nearest neighbors, are realized through intermediate O$^{2-}$ 
		$p$-orbitals. The hatched circle indicates a possible projection of an external dopant onto this plane. Inset: small deviations of two kinds of spins from antiparallelism 
		in the $xz$-plane due to weak Dzyaloshinskii-Moriya exchange.}
	\lb{fig1}
\end{figure}
The relevant structure for superconductivity in perovskite cuprates is the CuO$_{2}$ plane. In the absence of doping, all the oxygens are in the O$^{2-}$ ionization state 
with fully occupied $p$-shell and all the coppers are in the Cu$^{2+}$ state with a single non-compensated $d$-electron spin (see figure~\ref{fig1}), while the hole 
charge carriers by doping are expected to reside on the cation sites changing their state for Cu$^{3+}$ \cite{lokt}. Then, the spin and charge dynamics of these 
$d$-electrons are supposed to be mostly responsible for the unordinary physical properties of cuprate materials, in conformity with various known approaches 
\cite{plak,dell}, but the main accent of the present work is on a considerable restructuring of both types of excitation spectra under random perturbations from 
dopants. 

\subsection{Spin dynamics}
\lb{spin} For treating the spin order and spin excitations in the initial undoped structure, we use the simplest model Hamiltonian involving the strong AFM 
superexchange $J$ between nearest neighbor Cu spins (to impose their antiparallel orientation) as well as the weak terms of tetragonal anisotropy $\D J_\textrm{t}$ and of 
Dzyaloshinskii-Moriya antisymmetric exchange $D$ (to provide stability of the AFM order): 
\be
H_s = \sum_{\bn,\boldsymbol \d} \left[J\bs_\bn\cdot\bs_{\bn + \boldsymbol \d} - \D J_\textrm{t} s_\bn^y s_{\bn + \boldsymbol \d}^y - 
D\left(s_\bn^y s_{\bn + \boldsymbol \d}^z - s_\bn^z s_{\bn + \boldsymbol \d}^y\right)\right],
\lb{Hs}
\ee
with the vectors $\boldsymbol{\d}$ of separation between nearest neighbor Cu sites. 

Under the strongest $J$ effect only, the spin ground state would have two enclosed square sublattices with strictly opposite spin orientations but unstable because 
of their free absolute orientation. Adding the weak $\D J_\textrm{t}$ effect, restricts this freedom to the $xz$-plane and, finally, the weak $D$ effect fixes the sublattice quantization 
axes almost antiparallel along the $z$-axis and with slight symmetric deviations from $z$- towards $y$-axis by an angle $\theta$ (see inset in figure~\ref{fig1}). This 
implies the following expressions of spin components on $j$-th sublattice sites in $\bn$-th unit cell through the local spin-flip operators $b_{\bn_j}$: 
\bea
s_{\bn_1}^z & = & \left(\frac{1}{2} - b_{\bn_1}^\dagger b_{\bn_1}\right)\cos\theta + \frac{b_{\bn_1} - b_{\bn_1}^\dagger}{2\ri}\sin\theta, 
\qquad s_{\bn_2}^z = \left(b_{\bn_2}^\dagger b_{\bn_2} - \frac{1}{2}\right)\cos\theta + \frac{b_{\bn_2}- b_{\bn_2}^\dagger }{2\ri}\sin\theta,\nn\\
s_{\bn_1}^y & = & \left(b_{\bn_1}^\dagger b_{\bn_1} - \frac{1}{2}\right)\sin\theta + \frac{b_{\bn_1} - b_{\bn_1}^\dagger}{2\ri}\cos\theta,
\quad \quad s_{\bn_2}^y =  \left(b_{\bn_2}^\dagger b_{\bn_2} - \frac{1}{2}\right)\sin\theta - \frac{b_{\bn_2} - b_{\bn_2}^\dagger}{2\ri}\cos\theta,\nn\\
s_{\bn_1}^x & = & \frac{b_{\bn_1} + b_{\bn_1}^\dagger}2\,,\qquad\qquad\qquad\qquad\qquad\qquad \,\,\,\, s_{\bn_2}^x = \frac{b_{\bn_2} + 
	b_{\bn_2}^\dagger}2\,.
\lb{sf}
\eea
Being inserted in equation~\eqref{Hs}, they generate the terms up to fourth order in $b$-operators  but the ground state stability needs the linear terms to vanish (together with 
the third order terms) and this requires the deviation angle to be:
\be
\theta = \frac12\arctan\frac{2D}{2J - \D J_\textrm{t}} \ll 1.
\lb{theta}
\ee
Next, the quartic terms can be neglected for small excitation numbers, $\langle b_{\bn_j}^\dagger b_{\bn_j}\rangle \ll 1$, when $b_{\bn_{j}}$ can be seen as Bose 
operators (the Holstein-Primakoff ones) and the spin Hamiltonian turns bilinear in their Fourier-transforms $b_{j,\bk} = N^{-1/2}\sum_{\bn} \re^{\ri\bk\bn_{j}}
b_{\bn_{j}}$ (summed over $N$ unit cells):
\bea
H_s & = & \sum_\bk \left[A\left(b_{1,\bk}^{\dagger}b_{1,\bk} + b_{2,\bk}^{\dagger}b_{2,\bk}\right) + B\g_\bk\left(b_{1,\bk}
b_{2,-\bk} + b_{1,\bk}^\dagger b_{2,-\bk}^\dagger\right) + C\g_\bk\left(b_{1,\bk}^\dagger b_{2,\bk} + b_{2,\bk}^\dagger 
b_{1,\bk}\right) \right],\nn\\
A & = & 2J \cos 2\theta + 2\D J_\textrm{t}\sin^2\theta + 2D\sin2\theta,\quad B = J(1 + \cos 2\theta) - \D J_\textrm{t}\cos^2\theta + D\sin2\theta,\nn\\
C & = & J(1 - \cos 2\theta) + \D J_\textrm{t}\cos^2\theta - D\sin2\theta,\quad \g_\bk = \cos \frac{k_x}2 \cos \frac{k_z}2
\lb{Hsb}
\eea
(with wave-vectors $\bk$ measured in units of inverse lattice parameter $a^{-1}$). We notice that the difference between the big terms $A \approx 2J + D^2/J$ and 
$B \approx 2J - \D J_\textrm{t} + D^2/2J$ is of the order of the small term $C \approx  \D J_\textrm{t} - D^2/2J$.

The Hamiltonian, equation~\eqref{Hsb}, is readily diagonalized by a standard unitary transformation of the sublattice $b_{j,\bk}$ operators into the $\b_{\pm,\bk}$ operators of 
eigen-modes (symmetric or antisymmetric in the sublattice indices). This can be compactly expressed in terms of 4-spinors transformed by a 4$\times$4  matrix as 
$b_\bk = \hat U_\bk \b_\bk$, or in their components:
\be
b_\bk = \left(\begin{array}{c}
	b_{1,\bk} \\
	b_{1,-\bk}^\dagger\\
	b_{2,\bk} \\
	b_{2,-\bk}^\dagger  
\end{array}\right),\quad 
\hat U_\bk = \frac 1{\sqrt 2}\left(
\begin{array}{cccc}
	\cosh\theta_{+,\bk} &  \sinh\h_{+,\bk} & \cosh\h_{-,\bk}  &  -\sinh\h_{-,\bk} \\
	\sinh\h_{+,\bk} & \cosh\h_{+,\bk}   & -\sinh\h_{-,\bk}  & \cosh\h_{-,\bk}\\
	\cosh\h_{+,\bk}  &  \sinh\h_{+,\bk}  & -\cosh\h_{-,\bk} &  \sinh\h_{-,\bk}   \\
	\sinh\h_{+,\bk}  & \cosh\h_{+,\bk}  & \sinh\h_{-,\bk} &  - \cosh\h_{-,\bk} 
\end{array}
\right),\quad 
\b_\bk = \left(\begin{array}{c}
	\b_{+,\bk} \\
	\b_{+,-\bk}^\dagger\\
	\b_{-,\bk} \\
	\b_{-,-\bk}^\dagger  
\end{array}\right).
\lb{Usc}
\ee
Here, the 4$\times$4 matrices can be expressed as direct products of two kinds of Pauli matrices, $\hat \s_j$ in spin indices (e.g., between $b_{j,\bk}$ and 
$b_{j,\bk}^\dagger$) and $\hat \t_j$ in sublattice/mode indices (e.g., between $b_{1,\bk}$ and $b_{2,\bk}$). So, the above $U$-matrix is presented in this way 
as: 
\[\hat U_\bk = \left[\re^{\h_{+,\bk}\hat \s_1}\otimes\left( \hat \t_1 - \ri\hat \t_2 + \hat \t_3 + 1\right) + \re^{-\h_{-,\bk}\hat \s_1}\otimes
\left(\hat \t_1 + \ri\hat \t_2  + \hat \t_3 - 1\right)\right]/\sqrt8\,.\]  
This permits to reduce multiplication of 4$\times$4 matrices to that of 2$\times$2 ones by the rule: $[f(\hat\s_j)\otimes g(\hat\t_k)][f'(\hat\s_l)\otimes 
g'(\hat\t_m)] = [f(\hat\s_j)f'(\hat\s_l)]\otimes[g(\hat\t_k)g'(\hat\t_m)]$, for any functions $f$, $f'$, $g$, and $g'$.

Then, with the choice of hyperbolic rotation arguments as:
\be
\h_{\pm,\bk} = \frac 12 \arctanh\frac{B\g_\bk}{A \pm C\g_\bk}\,, 
\lb{uv1}
\ee
the diagonal $H_s$ form is reached:
\be
H_s =  \sum_{j = \pm}\sum_\bk \o_{j,\bk}\b_{j,\bk}^{\dagger}\b_{j,\bk}\,, \quad \textrm{with}\quad \o_{\pm,\bk} = \sqrt{\left(A \pm C\g_\bk\right)^2 
	- B^2\g_\bk^2}\,.
\lb{Hsd}
\ee
In the most important low-energy range, where $\g_\bk^2 \approx 1 - k^2/4$, the dispersion laws are approximated as:
\be
\o_{j,\bk} \approx \sqrt{\o_j^2 + J^2 k^2}\,,
\lb{sdl}
\ee
with the spectrum gaps $\o_- = 2D$ and $\o_+ = 2\sqrt{2J\D J_\textrm{t}}$. These values just testify that the rigidity of the antisymmetric mode (in-plane rotations generated 
mainly by the spin $x$-components) comes from the in-plane $D$ anisotropy and that of the symmetric mode (out-of-plane breathings by the $y$-components) does 
from the out-of-plane $\D J_\textrm{t}$ anisotropy. The spin dynamics of this spin system is contained in the GF matrix $\hat G_\bk  = \lGF \b_\bk|\b_{\bk}^\dagger\rGF$. In 
particular, DOS of spin excitations is defined as: 
\be
\rho(\o) = \frac 1{\piup N} \Im \Tr \sum_\bk  \hat G_\bk\,,
\lb{DOS1}
\ee 
where the sum over 2D Brillouin zone, as usually, can be substituted by integration according to the rule:
\be
\frac1N\sum_\bk f(\bk) = \frac1{4\piup^2} \int_{-\piup}^{\piup}\rd k_x  \int_{-\piup}^{\piup}\rd k_z f(\bk).
\lb{int}
\ee
For the unperturbed Hamiltonian, equation~\eqref{Hsd}, the GF matrix takes the simplest diagonal form, $\hat G_\bk \Rightarrow \hat g_\bk$, with its non-zero elements: 
\[\lGF \b_{j,\bk}|\b_{j,\bk}^\dagger\rGF = \frac 1{\o - \o_{j,\bk}}\,,\qquad \lGF \b_{j,\bk}^\dagger|\b_{j,\bk}\rGF = -\frac1{\o + \o_{j,\bk}}.\] 
Then, using the equation~\eqref{sdl}  provides a simple analytic approximation for the relevant integrals (also called the locator functions) in the low-energy range:
\be 
g_j(\o) = N^{-1}\sum_\bk \frac 1{\o - \o_{j,\bk}} \approx -\frac 1J \left(1 + \frac\o {2J}\ln\frac{2J}{\o_j - \o}\right),
\lb{andos}
\ee
and hence the low-energy approximation for DOS:
\be
\rho(\o) \approx \frac\o{4J^2}\left[\theta(\o - \o_-) + \theta(\o - \o_+)\right].
\lb{DOSa}
\ee

The AFM order stability is controlled by the  spectrum gaps $\o_\pm$, limiting thermal occupation of excited states over the fully ordered ground state. Eventually, 
at reaching  the N\'eel  temperature $T_{\text{N}}$, the mean sublattice magnetization, as, e.g., 
\[s = \langle s_{\bn_1}^z\rangle = \frac 12  -  \frac1N\sum_\bk \langle   b_{1,\bk}^\dagger b_{1,\bk}\rangle\,,\] 
should vanish. This average is expressed with the use of the spectral theorem, equation~\eqref{spt}, and the unitary transformation, equation~\eqref{Usc}, as:
\begin{eqnarray}
\frac 1N\sum_\bk \langle b_{1,\bk}^\dagger b_{1,\bk}\rangle &\approx& \frac 1{2N}\sum_{j,\bk} \left(\cosh^2\h_{j,\bk}\langle \b_{j,\bk}^\dagger \b_{j,\bk}\rangle 
+ \sinh^2\h_{j,\bk}\langle \b_{j,\bk}\b_{j,\bk}^\dagger \rangle\right) \nonumber \\ &\approx&  \frac {J}N\sum_{j,\bk} \frac{2n_{\text{B}}\left(\o_{j,\bk}\right) + 1}{\o_{j,\bk}} - \frac12\,,
\lb{sav}
\end{eqnarray}
involving the Bose occupation function: $n_{\text{B}}(\o) = (\re^{\b\o} -1)^{-1}$. Since the temperature dependent part of the sum in equation~\eqref{sav} is mainly 
due to the low-energy spectrum, this can be approximated by equation~\eqref{sdl} and the radial integration limit can be safely extended to infinity, giving for $j$th subband:
\[\frac {2J}N \sum_\bk \frac{n_{\text{B}}\left( \o_{j,\bk}\right)}{\o_{j,\bk}} \approx \frac{J }{\piup}\int_0^\infty\frac{n_{\text{B}}\left(\sqrt{\o_j^2 + J^2 k^2}\right)k \rd k }{\sqrt{\o_j^2 + 
		J^2 k^2}} = \frac 1{\piup J} \int_{\o_j}^\infty n_{\text{B}}(\o) \rd\omega = \frac {\ln\left[1 + n_{\text{B}}(\o_j)\right]}{\piup\b J}.\]
For temperatures much higher of the bandgaps, $\b\o_j \ll 1$, the latter logarithm is further simplified just to $\ln\left[1/(\b\o_j)\right]$. 

The temperature independent part in equation~\eqref{sav} (presenting spin zero-point fluctuations) is contributed by the whole spectrum by equation~\eqref{Hsd} and its numeric value 
$\a = N^{-1}\sum_{j,\bk} J/\o_{j,\bk} - 1/2 \approx 0.19$ is almost insensitive to the small parameters $\D J_\textrm{t}/J$ and $D/J$. Then, the N\'eel temperature is found from 
the equation:
\be
T_{\text{N}} \approx \frac{(1/2 - \a) \piup J}{k_{\text{B}}\ln\left[T_{\text{N}}^2/(\o_+\o_-)\right]}\,,
\lb{TN}
\ee
and, for the choice of parameters: $J = 135$ meV (as measured in La$_2$CuO$_4$ \cite{hay} and supported by  \textit{ab initio} calculations \cite{esk,bel} for all CuO$_2$ 
planes) and $\o_-\o_+ = 6.5$ meV$^2$, we have from equation~\eqref{TN}  this temperature as $T_{\text{N}} \approx 320$ K, in a good agreement with its experimental value for 
undoped La$_2$CuO$_4$ (and actually much higher of the chosen geometric average of bandgaps $\sqrt{\o_+ \o_-}/k_{\text{B}} \approx 30$ K). 

It can be yet noted that the stabilizing anisotropy $D$ factor for the AFM order may also remain important beyond this order range, for instance, defining the exotic 
pseudogap feature in the electronic spectrum near its Fermi level in the temperature range from $T_{\text{N}}$ to the much higher structural tetragonal-orthorhombic 
transition $T_\text{t{-}o}$ \cite{laugh,mori}, but this issue is left out of the present consideration.

\subsection{Charge dynamics}
\lb{charge} Another kind of dynamics results from much faster charge carrier hopping over Cu sites. Aiming to define the Fermi level position from the resulting 
spectrum, we start with the model hopping Hamiltonian referred to the atomic Cu $d$-term:
\be
H_{\text{hop}} = -\sum_{\bk,\s} \left[t_\bk (a_{1,\s,\bk}^\dagger a_{2,\s,\bk} + a_{2,\s,\bk}^\dagger a_{1,\s,\bk}) + t'_\bk (a_{1,\s,\bk}^\dagger  
a_{1,\s,\bk}  + a_{2,\s,\bk}^\dagger a_{2,\s,\bk})\right].
\lb{Hhop}
\ee
Here,  $a_{j,\s,\bk} = N^{-1/2}\sum_\bn \re^{\ri\bk\bn_j} a_{\bn_j,\s}$ are the Fourier transforms of the Fermi operators $a_{\bn,\s}$ for an electron with the given spin 
value $\s$ on $\bn$-th site. The hopping factors $t_\bk = 4t\cos \tfrac{k_x}2\cos \tfrac{k_z}2$ (for nearest neighbors) and $t'_\bk = 2t'(\cos k_x + \cos k_z)$ (next nearest 
neighbors) involve the hopping parameters $t$ and $t'$ (both taken with negative sign for their indirect realization through the excited O levels \cite{lokt}).  This model 
can be seen as a simplified version of the much more detailed LDA Hamiltonian \cite{anders}. 

Actually, the spin operators (in neglect of small deviations $\theta$ from $z$-axis, inessential for the charge dynamics) can be also presented in terms of the Fermi 
operators as:
\be
s_\bn^z = \tfrac12\left(a_{\bn,\ua}^\dagger a_{\bn,\ua} - a_{\bn,\da}^\dagger a_{\bn,\da}\right),\qquad s_\bn^+ = a_{\bn,\ua}^\dagger 
a_{\bn,\da}\,, \qquad s_\bn^- = a_{\bn,\da}^\dagger a_{\bn,\ua}\,.
\lb{sfo}
\ee
Then, the dominant purely exchange part of the spin Hamiltonian, equation~\eqref{Hs}, obtains in these terms a four-fermion structure:
\bea
H_\text{ex} & = & J \sum_{\bn,\boldsymbol \d} \left[\tfrac 14 \left(a_{\bn_1,\ua}^\dagger a_{\bn_1,\ua} - a_{\bn_1,\da}^\dagger a_{\bn_1,\da}\right)
\left(a_{\bn_1 + \boldsymbol \d,\ua}^\dagger a_{\bn_1 + \boldsymbol \d,\ua} - a_{\bn_1 + \boldsymbol \d,\da}^\dagger a_{\bn_1 + 
	\boldsymbol \d,\da}\right)\right.\nn\\
& + & \left. \tfrac 12\left(a_{\bn_1,\ua}^\dagger a_{\bn_1,\da} a_{\bn_1 + \boldsymbol \d,\da}^\dagger a_{\bn_1 + \boldsymbol \d,\ua} + 
a_{\bn_1,\da}^\dagger a_{\bn_1,\ua} a_{\bn_1 + \boldsymbol \d,\ua}^\dagger a_{\bn_1 + \boldsymbol \d,\da}\right)\right],
\lb{H4f}
\eea
or, passing to the band operators:
\bea
H_\text{ex} & = & \frac J N \sum_{\bk,\bk',\bq} \left[\left(a_{1,\ua,\bk}^\dagger a_{1,\ua,\bk'} - a_{1,\da,\bk}^\dagger a_{1,\da,\bk'}\right)
\left(a_{2,\ua,\bk + \bq}^\dagger a_{2,\ua,\bk' - \bq} - a_{2,\da,\bk + \bq}^\dagger a_{2,\da,\bk' - \bq}\right)\right.\nn\\
& + & \left. 2\g_{\bk - \bk'}\left(a_{1,\ua,\bk}^\dagger a_{1,\da,\bk'} a_{2,\da,\bq - \bk}^\dagger a_{2,\ua,\bq - \bk'} + 
a_{1,\da,\bk}^\dagger a_{1,\ua,\bk'} a_{2,\ua,\bq - \bk}^\dagger a_{2,\da,\bq - \bk'}\right)\right].
\lb{H4b}
\eea
This can be also reduced to a bilinear form, like that in equation~\eqref{Hhop}, by a common separation of some non-zero pair averages beside the resting operator pair 
products (and neglecting as usually small fluctuations beside the averages).  We note that this known procedure involves here either ``normal'', $\langle a^\dagger 
a\rangle$-type, and ``anomalous'', $\langle a a\rangle$- or $\langle a^\dagger a^\dagger\rangle$-type averages. They will express respectively the spin exchange 
effects on the normal electronic spectrum and on its SC restructuring.

Particularly, the ``normal'' averages relate to the local spin values and they are selected from the first, longitudinal part under the sum in equation~\eqref{H4f} as:
\be
\langle a_{\bn_1,\ua}^\dagger a_{\bn_1,\ua} - a_{\bn_1,\da}^\dagger a_{\bn_1,\da}\rangle = - \langle a_{\bn_2,\ua}^\dagger 
a_{\bn_2,\ua} - a_{\bn_2,\da}^\dagger a_{\bn_2,\da}\rangle = 2s,
\lb{nav}
\ee
where the $s$ value can be evaluated (in the function of temperature and doping) from the above considered Bose-type GFs. Otherwise, the ``anomalous'' averages 
are due to delocalized pairs of electrons with opposite spins and momenta and they follow from both parts in equation~\eqref{H4b} (to be detailed later).

Now, we compose the effective two-fermion Hamiltonian for normal electrons: $H_\text{n} = H_{\text{hop}} + H_\text{n}^{(2)}$, involving the ``normally'' averaged spin exchange part:
\be
H_\text{n}^{(2)} = J \sum_\bk \left(a_{1,\da,\bk}^\dagger a_{1,\da,\bk} + a_{2,\ua,\bk}^\dagger a_{2,\ua,\bk} - a_{1,\ua,\bk}^\dagger  a_{1,\ua,\bk} 
- a_{2,\da,\bk}^\dagger a_{2,\da,\bk}\right),
\lb{Hn2}
\ee
which can be seen as a spin-Hubbard repulsion, favouring an electron with the given spin to occupy the sites from its proper sublattice (1 for $\ua$ and 2 for $\da$).

Similarly to the above boson case,  we construct fermion 4-spinors in sublattice/spin indices: $a_\bk^\dagger = (a_{1,\ua,\bk}^\dagger,a_{2,\ua,\bk}^\dagger,
a_{2,\da,\bk}^\dagger,a_{1,\da,\bk}^\dagger)$, and transform them into such spinors in mode/spin indices: $\a_\bk^\dagger = (\a_{+,\ua,\bk}^\dagger,\a_{-,\ua,\bk}^\dagger,
\a_{+,\da,\bk}^\dagger,\a_{-,\da,\bk}^\dagger)$ by a unitary rotation:
\be
a_\bk = \re^{\ri\l_\bk\hat\t_2}\otimes \hat 1 \a_\bk\,,
\lb{Uhop}
\ee
(with the direct product of $\hat\t$-matrices in mode indices and unit matrix in spin indices). Then, the choice of this rotation argument $\l_\bk = \tfrac12\arctan t_\bk/J$ 
brings the effective two-fermion Hamiltonian to its diagonal form:
\be
H_\text{n} = \sum_{j,\s,\bk} \e_{j,\bk} \a_{j,\s,\bk}^\dagger \a_{j,\s,\bk}\,,
\lb{Hn}
\ee
with 4 composite eigen-modes, of spin-majority ($j = +$) or spin-minority ($j = -$) composition [not to be mixed up with the symmetric/antisymmetric boson modes 
in equation~\eqref{Hsd}] for each $\s$ value. The corresponding eigen-energies:
\be
\e_{\pm,\bk} = t'_\bk \mp \sqrt{J^2 + t_\bk^2}\,,
\lb{majmind}
\ee
(note their double degeneracy in spin index) display the possibility for an energy gap to open between the top of the majority subband, $\e_{+,\text{max}} = 4t' - J$ reached 
at the BZ M-points, $\bk = (\pm\piup,\pm\piup)$, and the bottom of minority bands, $\e_{-,\text{min}} = J$ at the X-points, $\bk = (0,\pm\piup)$, or Y-points, $\bk = (\pm\piup,0)$. This 
most important spin-exchange effect appears under the condition $J \geqslant 2t'$, realistic for strong AFM coupling $J$ in cuprates. In the absence of doping, it defines the 
system insulating state with fully occupied majority bands and empty minority bands, the Fermi energy (the chemical potential at $T = 0$) being located at the gap 
middle point: $\m = 2t'$. In this Fermi energy reference, the subband edges are written as: $\e_{-,\text{min}} = J - 2t' = -\e_{+,\text{max}}$. More details of this spectrum are seen 
from its DOS function:
\be
\r(\e) = \frac 1{\piup N} \sum_{j,\s,\bk} \Im \lGF\a_{j,\s,\bk}|\a^\dagger_{j,\s,\bk}\rGF = \frac 2{N} \sum_{j,\bk} \d\left(\e - \e_{j,\bk}\right),
\lb{DOS}
\ee
where the corresponding integration can be simplified, passing from the variables $k_x$ and $k_z$ to $u = \cos \tfrac{k_x}2 \cos \tfrac{k_z}2$ and $v = 1 + (\cos k_x + 
\cos k_z)/2$ and using the Jacobian:
\[\frac{{\mathcal D}(u,v)}{{\mathcal D}\left(k_x, k_y\right)} = \frac14\sin \tfrac{k_x}2 \sin \tfrac{k_z}2\left(\cos k_z - \cos k_x\right).\]
The latter is then expressed in the function of $u,v$:
\[\frac{{\mathcal D}(u,v)}{{\mathcal D}\left(k_x, k_z\right)} = - \frac12\sqrt{\left(u^2 + 1 - v\right)\left(v^2 - 4u^2\right)}\,.\]
Also conversion of the DOS integrals by equation~\eqref{DOS} to the $u,v$ variables requires to express the dispersion laws as
\[\e_{\pm,\bk} \equiv \e_\pm(u,v) = -4t'v \mp \sqrt{J^2 + 16t^2u^2}\]
and to modify the integration limits. 
\begin{figure}[!b]
	\center \includegraphics[width=10cm]{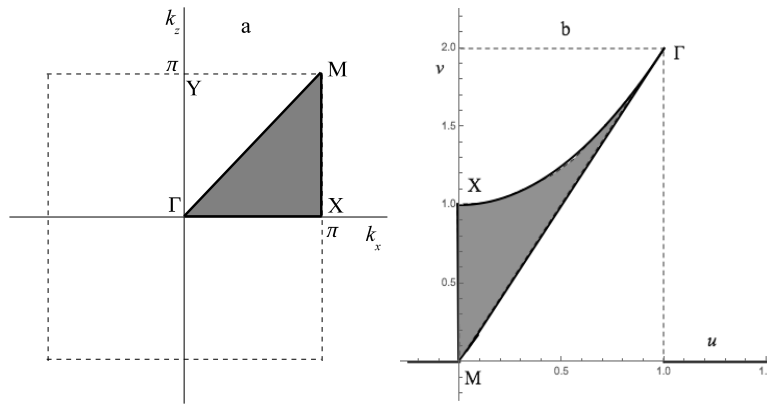}\\
	\caption{a) The BZ relevant integration area (shadowed) in the $k_x, k_z$ variables (in $a^{-1}$ units). b) The same in the $u,v$-variables.}
	\lb{fig2}
\end{figure}

The relevant integration area between BZ points $\G$, X, and M is delimited in the $k_x,k_z$ variables [figure~\ref{fig2}(a)] by the lines: $k_z = 0$ at $0 \leqslant k_x \leqslant \piup$ 
($\G$-X), $k_z = k_x$ at $0 \leqslant k_x \leqslant \piup$ ($\G$-M), and $k_x = \piup$ at $0 \leqslant k_z \leqslant \piup$ (X-M). In the $u,v$ variables [figure~\ref{fig2}(b)] they are converted to the 
lines:  $u = 0$ at $0 \leqslant v \leqslant 1$ (M-X),  $v= 2u$ at $0 \leqslant u \leqslant 1$ (M-$\G$), and $v = 1 + u^2$ at $0 \leqslant u \leqslant 1$ (X-$\G$). Eventually, the DOS integrals in these 
variables are written as:
\[\r_\pm(\e) = \frac1{2\piup^2}\int_0^1 \rd u\int_{2u}^{1 + u^2}\rd v\frac{\d\left(\e - 4t'v \pm \sqrt{J^2 + 16t^2 u^2}\right)}{\sqrt{\left(u^2 + 1 - v\right)\left(v^2 - 4u^2\right)}},\]
and, after the trivial $\d$-function integration in $v$, they are reduced to one-dimensional $u$-integrals:
\be
\r_\pm(\e) = \frac 1{8\piup^2 |t'|} \int_0^1  \frac{\theta\left(u^2 + 1 - v_\pm\right)\theta\left(v_\pm - 2u\right)}{\sqrt{\left(u^2 + 1 - v_\pm\right)\left(v_\pm^2 - 4u^2\right)}}\rd u\,,
\lb{dens}
\ee
with $v_\pm(\e,u) = (\e  \pm \sqrt{J^2 + 16t^2u^2})/4t'$. Then, an easy numerical calculation results in the DOS profiles shown in figure~\ref{fig3} for a realistic choice of 
the hopping parameters: $t = 0.5$ eV and $t'= 50$ meV, compared to the above chosen exchange parameter $J = 135$ meV. They present two types of van Hove 
singularities: 

\begin{figure}[!t]
\center \includegraphics[width=0.6\textwidth]{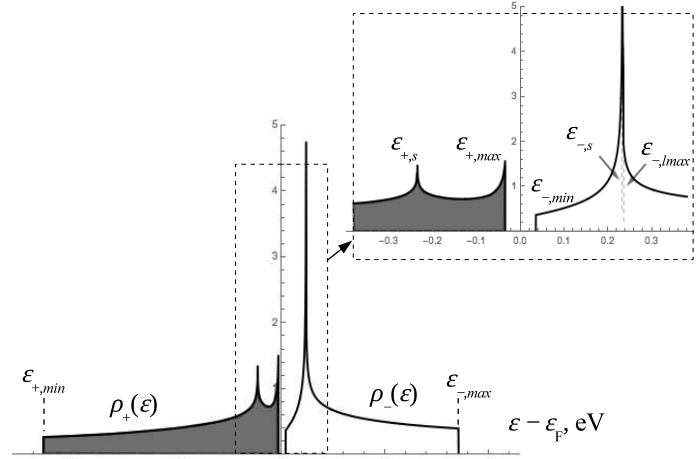}
	\caption{Density of states with the energy gap between spin-majority (filled) and spin-minority (empty) electronic subbands and the van Hove singularities at their 
		characteristic energies (see the text) calculated from equation~\eqref{dens} at the choice of the model parameters: $t = 0.5$ eV, $t'= 50$ meV, $J = 135$ meV. Its more 
		detailed behavior in the gap range (delimited by the dashed rectangle) is shown in the inset.}
	\lb{fig3}
\end{figure}

1) discontinuous steps at the subband edges, $\e_{+,\text{min}} = -6t'- \sqrt{J^2 + 16t^2}$, $\e_{+,\text{max}} = 2t' - J$, $\e_{-,\text{min}} = J$, $\e_{-,\text{max}} = 2t' + \sqrt{J^2 + 16t^2}$, 
and also at the minority subband local maximum $\e_{-,\text{lmax}} = 2t' + J$, 

2) logarithmic peaks at the saddle points, $\e_{+,s}= - 2t' - J$ and $\e_{-,s} = 2t' + J\sqrt{1 - (t'/t)^2}$.

A notable asymmetry between the majority and minority subbands seen in figure~\ref{fig3}, in particular, in their bandwidth and edge DOS levels, can serve for understanding  the experimentally detected difference in the cuprate behavior under hole and electron types of doping \cite{dell}, but we restrict the present consideration only to the 
more habitual hole type. 

In the following analysis of the spectrum restructuring under such doping, into metallic and further into SC state, a special importance pertains to the $\e_{+,\bk}$ 
isoenergetic lines in BZ around its M-point that display $C_4$ symmetry as shown in figure~\ref{fig4}, in general similarity to the known experimental measurements of these 
lines with the ARPES techniques \cite{kord}. Their maximum distance $q_{\text{max}}$ from the M-point (along the $k_x$ and $k_z$ axes) is written in the function of the energy 
depth $\d = \e_{+,\text{max}} - \e$ as:
\be
q_{\text{max}} = 2\arcsin \sqrt{\frac{\d}{4t'}}\,,
\lb{qmax}
\ee
and the minimum distance (along the $k_x = \pm k_z$ lines) does as:
\be
q_{\text{min}} = 2\sqrt2\arcsin \sqrt{\frac{\sqrt{t^2\d(2J + \d) + (2Jt')^2} - 2t'(J + \d)}{4(t^2 - 4t'^2)}}\,,
\lb{qmin}
\ee
but in the most relevant energy range, $J(t'/t)^2 \ll \d \ll J$, the latter is well approximated as: 
\be
q_{\text{min}} \approx \sqrt8 \arcsin \sqrt[4]{\frac{\d}\O}\,,
\lb{qmaxmin}
\ee
with a much greater energy scale $\O = 8\left(t - 4t'^2/t\right)^2/J \gg J$.
Then, the approximated dispersion law for $\d$ in polar coordinates around the M-point can be presented as:
\be
\d(q,\varphi) \approx 4t' \sin^2\frac q2 \cos^2\varphi + \O\sin^4\frac q{\sqrt8} \sin^2\varphi.
\lb{qphi}
\ee
Such factorization of radial $q$- and azymuthal $\varphi$- dependencies essentially simplifies the further analysis of the system spectral characteristics 
under doping.
\begin{figure}[!t]
	\center \includegraphics[width=6cm]{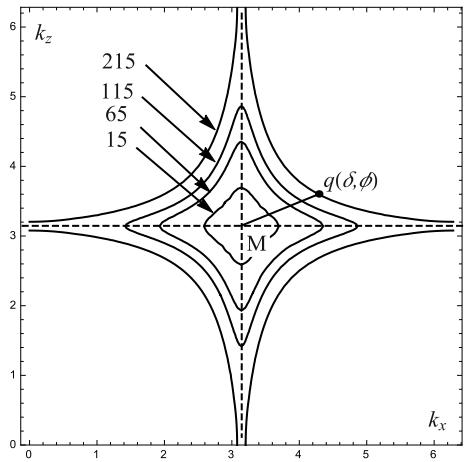}\\
	\caption{Isoenergetic lines $q(\d,\phi)$ for momentum relative to the M-point of BZ at different energy depths $\d$ below the top of the majority subband by equation~\eqref{majmind} (their values in meV indicated by the arrows) present the $C_4$ symmetry in the azymuthal angle $\phi$ and are transformed with growing $\d$ from 
		separate arcs to continuous lines around the BZ $\G$-point. The same symmetry and topology of isoenergetic lines is also preserved at finite doping describing the 
		evolution of Fermi lines and associated energy levels.}
	\lb{fig4}
\end{figure}

\section{Doping effects}
\lb{dop}
Now, we pass to the effects of doping on both kinds of excitation spectra. First of all, there are two principal perturbation mechanisms for a non-isovalent 
dopant atom (located out of the relevant CuO$_2$ plane) on the insulating host with the AFM spin order: 

1) the dopant Coulomb potential, acting on in-plane Cu sites within a certain screening radius $r_{\text{scr}}$, and

2) appearance of a charge carrier (a hole), localized by this potential within a certain localization radius $r_{\text{loc}}$. 

Generally, these mechanisms can influence all the parameters in the original Hamiltonians, equations~\eqref{Hs} and \eqref{Hhop}, producing quasiparticle scattering 
of rather complicated structure and considerable technical problems for its analysis. 

Nevertheless, a qualitatively reasonable physical picture for such a disordered system is expected to follow already from the simplest one-parametric 
Lifshitz perturbation models for each diagonalized Hamiltonian, equations~\eqref{Hsd} or \eqref{Hn}. Of course, strictly speaking, the Lifshitz model with a single 
momentum-independent scattering amplitude should correspond to a purely point-like perturbation potential in the real space, or, in more general, to a 
perturbation that affects a single local component of a plane-wave excitation state (that is, its inverse Fourier transform). However, in the present system, the 
$\a_\bk$ and $\b_\bk$ eigen-modes are composed of the original $a_\bk$ and $b_\bk$ modes, according to equations~\eqref{Uhop} and \eqref{Usc},  and their 
inverse Fourier transforms:
\be
\a_\bn = \frac1{\sqrt N}\sum_\bk \re^{-\ri\bk\bn}\a_\bk = \sum_{\bn'}\hat f^{(a)}(\bn - \bn')a_{\bn'}\,,\qquad \b_\bn = \frac1{\sqrt N}\sum_\bk \re^{-\ri\bk\bn}
\b_\bk = \sum_{\bn'}\hat f^{(b)}(\bn - \bn')b_{\bn'}\,,
\lb{FT}
\ee
are the distributions of the original point-like $a_\bn$ and $b_\bn$ with some kernel matrices $\hat f^{(a,b)}(\br)$. The latter matrix elements are made of 
the inverse Fourier transforms of $\cos \l_\bk$, $\sin \l_\bk$ (for $\a$-modes) or $\cosh \h_{\pm,\bk}$, $\sinh \h_{\pm,\bk}$ (for $\b$-modes) and they have as 
long decay range as $r^{(b)} \sim J/\o_j \gg 1$ for boson $\b$-states but only $r^{(a)} \sim t/J \sim 1$ for fermion $\a$-states. Therefore, the above mentioned 
perturbations of $\sim r_{\text{loc}}$ or $\sim r_{\text{scr}}$ range can be felt by the broadly distributed boson modes as effectively point-like and that may justify the Lifshitz 
model approximation for them. As to the sharper resolved fermion modes, where a more detailed perturbation model should be applied, fortunately it does not 
make essential technical problems to obtain an effective Lifshitz approximation also for this case (to be shown later). 

Now, beginning from the spin excitations, we supplement equation~\eqref{Hsd} with the Lifshitz perturbation Hamiltonian of the form:
\be
H'_s = - \frac {V_s}N \sum_{j,\bk,\bk',\bp} \re^{\ri(\bk - \bk')\bp} \b_{j,\bk}^\dagger \b_{j,\bk'}\,,
\lb{Hs1}
\ee
where $\bp$ are the  on-plane projections of random dopant positions (see figure~\ref{fig1}) and the negative sign of the single perturbation parameter $V_s$ 
indicates a general effective suppression of spin exchange and anisotropies due to the dopant potentials on Cu and O sites and to the reduction of spin density 
because of hole carriers. Then, in the standard general expression for GF matrix $\hat G_\bk = (\hat g_\bk^{-1} - \hat \S_\bk)^{-1}$, the simplest approximation 
for the self-energy matrix is: $\hat \S_\bk \approx c\hat T_s$,  where the momentum-independent (but energy-dependent) diagonal T-matrix for spin states:
\be
\hat  T_s(\o) =\left[\frac{1 + \hat\t_3}2 T_+(\o) + \frac{1 - \hat\t_3}2  T_-(\o)\right]\otimes\hat 1,
\lb{Tm}
\ee
and the subband terms $T_\pm(\o) = -V_s/\left[1 +V_s g_\pm(\o)\right]$ involve the locator functions by equation~\eqref{andos}. 

It is well known for disordered systems that zeroes of T-matrix denominators beyond the band spectrum define localized impurity levels in the diluted 
limit of doping and those can produce new impurity subbands at higher doping levels. Since each $g_j(\o)$ is negative and logarithmically diverging 
near the respective subband edge $\o_j$ , such levels $\o_{j,\text{loc}}$ are formally produced by both terms in equation~\eqref{Tm}, but of extremely shallow depth:
\be
\o_j - \o_{j,\text{loc}} \approx 2J \exp\left[-\frac{2J(J - V_s)}{\o_j V_s}\right],
\lb{omloc}
\ee
due to the huge $\sim J/\o_j \gg 1$ factor in the negative exponent (except for an improbable closeness of $V_s$ to $J$ within to $\sim \o_j$), so they are 
practically stuck to those edges. Moreover, the fast decay of logarithms in $g_\pm(\o)$ by equation~\eqref{andos} beyond an exponentially narrow vicinity of the edge, 
$\o_j - \o \sim J(\o_j - \o_{j,\text{loc}})/\o_j \ll \o_j$, makes them negligible besides the resting constant and then the only doping effect in the spin spectrum remains 
in equal downshifts of the subband edges as:
\be
\o_j(c) = \o_j - \frac{cV_sJ}{J - V_s}.
\lb{gapc}
\ee

This immediately defines the concentrational decay of the N\'eel temperature, $T_\text{N}(c)$, calculated again from equation~\eqref{TN} but with the change of $\o_j$ for 
$\o_j(c)$. In particular, the critical condition when  the smaller gap $\o_{\text{min}}(c) = \min_j\o_j(c)$ reaches zero [and so $T_\text{N}(c)$, with no more long-range 
AFM order in the system]  defines the critical doping value
\be
c_{\text{cr}} = \o_{\text{min}}(0)\left(\frac1{V_s}  - \frac1J\right).
\lb{ccr}
\ee
At $c < c_{\text{cr}}$, for a given product $\o_+\o_-$ [e.g., fitted to the known $T_\text{N} = T_\text{N}(0)$ as in section~\ref{spin}], the fastest decay of $T_\text{N}(c)$ is evidently attained 
for the case of equal gaps, $\o_+ = \o_- = \o_\pm$, and the corresponding behavior is presented in figure~\ref{fig5}, at the choice of perturbation parameter as 
$V_s = \o_\pm/(c_{\text{cr}} + \o_\pm/J)$. Notably, with the use of the experimental value of $c_{\text{cr}} \approx 0.02$ in La$_2$CuO$_4$, this parameter results $V_s = 65.5$~meV which is safely far from the adopted value of $J = 135$~meV, justifying the above general conclusions on the magnon spectrum reordering under doping. 
\begin{figure}[!t]
	\center \includegraphics[width=0.4\textwidth]{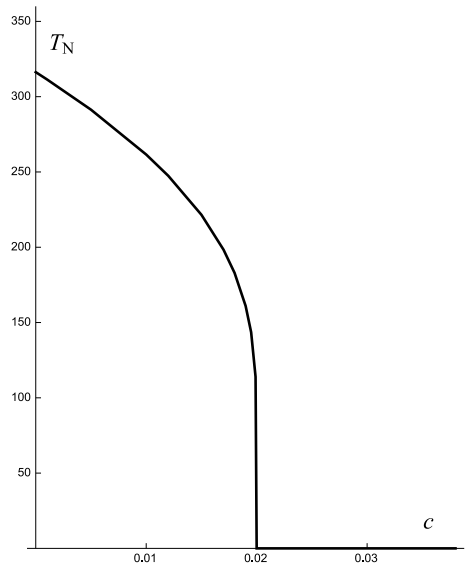}
	\caption{Doping dependence of the N\'eel temperature calculated from equation~\eqref{TN} at parameter values $\o_+ = \o_- = 2.55$ meV, $V_s = 65.5$ meV.}
	\lb{fig5}
\end{figure}

Now, passing to the charge carrier spectrum, we consider the perturbation to the Hamiltonian, equation~\eqref{Hn} from equal upshifts of atomic energy levels by $V_c$ 
on nearest neighbor Cu sites to a dopant (due to its repulsive potential for electrons). Then, in the presence of the AMF spin order (at least, local) in the lattice, there 
exist two types of dopant positions, with the up-spin neighbor sites along the $x$- or $z$-axes (the latter case just presented in figure~\ref{fig1}). The corresponding 
perturbation Hamiltonian takes the form:
\be
H'_{\text{hop}} = \frac1N\sum_{j=x,z}\sum_{\bp_j,\bk,\bk'} \re^{\ri(\bk - \bk')\bp}\a_\bk^\dagger\hat V^{(j)}_{\bk,\bk'}\a_{\bk'}\,,
\lb{Hh'}
\ee
with the scattering matrices:
\be
\hat V^{(x)}_{\bk,\bk'} = \frac{V_c}2\re^{\ri\l_\bk \hat\t_2}\left[\left(1 + \hat\t_3\right)\cos\frac{k_x - k'_x}2 + \left(1 -  \hat\t_3\right)\cos\frac{k_z - k'_z}2\right]
\re^{-\ri\l_{\bk'} 
	\hat\t_2}\,,
\lb{Vx}
\ee
and $\hat V^{(z)}_{\bk,\bk'}$ only differing from the above by the interchange of cosine factors at the $1 \pm \hat\t_3$ matrices. Then, within the T-matrix 
approximation, these two types of dopants give equal additive contributions to the carrier self-energy and we can do the further consideration only for one of 
them, e.g., for the $x$-type by equation~\eqref{Vx}, multiplying the result by the total number of dopants. 

In this course, we focus on the energy gap $\e_{+,\text{max}} \leqslant \e \leqslant \e_{-,\text{min}}$  between the subbands and specifically on the possibility for in-gap localized 
levels to appear here and to subsequently generate narrow impurity bands. The general observation is that, for positive $V_c$ value, such levels can be only 
created by the majority subband whose corresponding locator function $f_+(\e) = N^{-1}\sum_\bk(\e - \e_{+,\bk})^{-1}$ is positive within the gap, but not by the 
minority one with $f_-(\e) < 0$ here. So, the most important momentum range for the majority locator is the vicinity of M-point in BZ, where both scattering matrices 
tend to a simple constant: $\hat V^{(x,z)}_{\bk,\bk'} \approx V_c\left[1 + O(q_+^2)\right]$, thus justifying the Lifshitz approximation also for this case.

In the framework of this approximation, the T-matrix for carrier states is obtained in an analogy to the boson case, equation~\eqref{Tm}:
\be
\hat T_\text{c}(\e) = \left[\frac{1 + \hat\t_3}2 T_+(\e) + \frac{1 - \hat\t_3}2  T_-(\e)\right]\otimes\hat 1,
\lb{Tmc}
\ee
but with the subband terms  $T_\pm(\e) = V_c/\left[1 - V_cf_\pm(\e)\right]$. Here, the majority locator $f_+$ (namely, its relevant real part) can be well 
approximated using the expansion of equation~\eqref{qphi} within  its $\sim q^2$ term:
\be
f_+(\e) \approx \frac 1{4\piup t'}\ln\frac{4\piup t'}{\e - \e_{+,\text{max}}}
\lb{f+}
\ee
[to be compared with equation~\eqref{andos} for boson $g$-functions], and this readily defines the position of localized in-gap level:
\be
\e_{\text{loc}} = \e_{+,\text{max}} + 4\piup t'\re^{-4\piup t'/V_c},
\lb{eloc}
\ee 
whose separation from the band edge $\e_{+,\text{max}}$ occurs sizable compared to the band gap. Thus, choosing the perturbation parameter $V_c = 0.22$ eV 
and the previous parameters of the host system, we obtain this separation $\e_{\text{loc}} - \e_{+,\text{max}} \approx 36$ meV, nearly half of the gap, $\e_{-,\text{min}} - \e_{+,\text{max}} 
= 70$ meV. 

At finite dopant concentration $c$, the renormalized dispersion equation for the majority subband (in T-matrix approximation):
\be
\e - \e_{+,\bk} - c \Re T_+(\e) = 0\,,
\lb{deq}
\ee
describes a slightly modified main band and also a new impurity subband that arose near $\e_{\text{loc}}$, as seen in their modified DOS: $\tilde\r(\e) = \r_+[\e - c\Re
T_+(\e)]$ presented in figure~\ref{fig6}. This function permits to evaluate the necessary precondition for the onset of superconductivity in the doped system, its 
transition from original insulating to metallic state with the growing doping. First of all, this requires that the Fermi level, $\e_{\text{F}}$, (located within the bandgap in the
absence of doping) should occur within the continuous spectrum, and we can check this condition through analysis of the DOS in figure~\ref{fig6}. It is very important that 
the full weight of the impurity subband: 
\be
w_{\text{imp}} = \int_{\e_{\text{loc}}}^{\e_{\text{imp},\text{max}}}\tilde\r(\e)\rd\e\,,
\lb{wimp}
\ee
\begin{figure}[!b]
\vspace{4mm}
	\center \includegraphics[width=10.3cm]{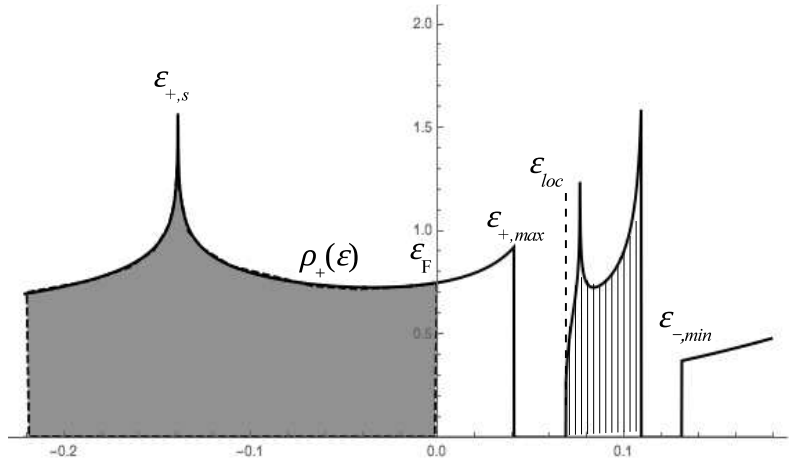}\\
	\caption{Density of carrier states near the band gap in presence of dopants with concentration $c = 0.1$ and perturbation potential $V_c = 0.22$ eV. The impurity 
		subband (hatched) formed above the localized impurity level $\e_{\text{loc}}$ has the weight of only $w_{\text{imp}} \approx 0.34c$ when the rest $0.66c$ of the doped holes 
		occupy the top of majority subbands (with $\s = \ua,\da$) above the Fermi level. Besides the overall spectrum shift to the $\e_{\text{F}}$ reference, its certain deformation 
		\textit{vs} the undoped case in figure~\ref{fig3} is seen, e.g., in the van Hove $\e_{+,s}$ peak getting closer to the band top $\e_{+,\text{max}}$ while the initial bandgap $(\e_{+,\text{max}},
		\e_{-,\text{min}})$ getting wider.}
	\lb{fig6}
\end{figure}is found notably smaller than the total weight $c$ of the doped holes (for instance, $w_{\text{imp}} \approx 0.34 c$ for the case in figure~\ref{fig6}), due to a certain weight transfer 
from the impurity subband to its generating majority subband. Thus, the resting $c - w_{\text{imp}}$ holes should occupy a certain range from $\e_{+,\text{max}}$ to $\e_{\text{F}}$ 
within the majority subband, according to:
\be
\int_{\e_{\text{F}}}^{\e_{+,\text{max}}}\tilde\r(\e)\rd\e = \frac{c - w_{\text{imp}}}2
\lb{eF}
\ee
(taking into account the spin degeneracy of the majority subband), and the numeric solution of this equation with respect to $\e_{\text{F}}$ gives the Fermi level depth 
$\d_{\text{F}}$ in function of doping $c$, shown by the solid line in figure~\ref{fig7}. From comparison with the profiles in figure~\ref{fig3}, this level defines the Fermi line in 
BZ as a set of disjoint Fermi arcs, in correspondence with the experimental observations in underdoped perovskites~\cite{doir}.
\begin{figure}[!t]
	\center \includegraphics[width=8cm]{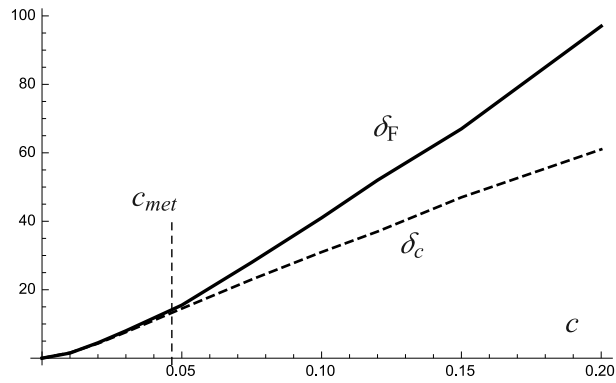}\\
	\caption{Doping dependent Fermi level depth $\d_{\text{F}}$, obtained from equations~\eqref{wimp}, \eqref{eF}, in comparison with that for the Mott mobility edge,  $\d_c$, estimated 
		from equation~\eqref{irm}.}
	\lb{fig7}
\end{figure}

However, the non-zero Fermi depth is not yet sufficient for the system metallization. It is important to take in mind that the dispersion equation, equation~\eqref{deq}, only makes 
sense for truly band-like states with well defined momentum values, satisfying the known Ioffe-Regel-Mott (IRM) criterion \cite{ir,mott}:
\be
\bq_+ \cdot{\boldsymbol \nabla} \e_{+,\bk} \gg  \G(\e)\,,
\lb{irm}
\ee
with the quasiparticle inverse lifetime $\G(\e) = c \Im T_+(\e)$. The energy range where this criterion fails is occupied by randomly localized states, not contributing 
either to metallic conductivity or to SC condensate, though the above estimates of their integrated DOS in equations~\eqref{wimp}, \eqref{eF} with $\tilde\r(\e)$ are not sensitive 
to this. The threshold between the band-like and localized ranges, the Mott mobility edge $\e_c$, is estimated from equation~\eqref{irm} at $\gg$ changed for $\sim$. Then, a
sufficient condition of metallization just consists in  the Fermi level occurring within the band-like range, $\e_{\text{F}} < \e_c$. 

Now, the straightforward use of equations~\eqref{dens}, \eqref{f+} in equation~\eqref{irm} gives an estimate for the mobility edge depth $\d_c$ in function of doping (the dashed line in figure~\ref{fig6}). Its comparison with the Fermi depth $\d_{\text{F}}$ function shows that the two should be very close until the doping value $\approx 0.05$ and then a growing dominance of 
$\d_{\text{F}}$ after this value, indicating it as the critical value $c_{\text{met}}$ for metallization in the considered model, in agreement with its experimental observation in doped 
perovskite La$_2$CuO$_4$.

Moreover, it should be noted that the effective weight of conductive states in such a metallic system results anomalously low:
\be 
w_{\text{met}}(c) = \int_{\e_c(c)}^{\e_{\text{F}}}\r(\e)\rd\e \ll c,
\lb{wc}
\ee
with respect to the total $c$ of the carriers doped into the system, as seen from the numerical calculation by equation~\eqref{wimp} shown in figure~\ref{fig8}, this deficit being at the 
cost of the impurity subband and of the localized range in the main subband. At last, there is a progressive deformation of the overall spectrum with doping, as seen in 
some compression of the upper part of the majority subband and in extension of the gap between the majority and minority subbands in figure~\ref{fig6} compared to figure~\ref{fig3}.
\begin{figure}[!t]
	\center \includegraphics[width=8cm]{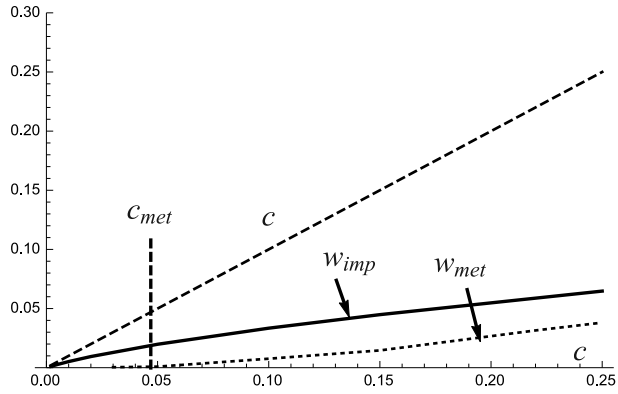}\\
	\caption{Doping dependence of impurity subband weight from equations~\eqref{wimp}, \eqref{eF}, and the mobile carriers weight $w_{\text{met}}$, compared to the total weight $c$ of the doped 
		carriers.}
	\lb{fig8}
\end{figure}

Now, we are in a position to discuss the onset and properties of SC state in the doped perovskite system with the above defined normal state properties.

\section{Superconductivity}
\lb{sc}
As it was already referred, the unusual features of SC state in doped perovskite compounds are mainly treated in literature within different scenarios of strongly 
correlated electrons  in a uniform medium, also incorporating the doping effects into this framework \cite{lee}. Here, we instead consider the narrow $J$-induced gap 
and explicit localization by dopant  potential as effective factors of such strong correlations. A similar approach was recently used for a description of destructive effects 
by magnetic impurities on high-$T_\text{c}$ SC order  \cite{gast}, \cite{lp1} but now we focus on its relevance for establishing of such an order. Namely, the possibility for SC 
coupling between charge carriers in the considered system is brought  by the ``anomalously''  averaged part of its four-fermion Hamiltonian, equation~\eqref{H4b}. Then, using 
the unitary transformation, equation~\eqref{Uhop}, the eigen-modes from the relevant  majority  subband (for brevity, this  subband index is omitted in what follows) are extracted 
from both longitudinal and transversal terms in equation~\eqref{H4b}, enabling the common formation  of Cooper  pairs. Thus, we get the  anomalous part of the SC Hamiltonian 
as:
\be
H_\text{an}^{(\text{sc})} =  \sum_\bk \left(\D_\bk \a_{\ua,\bk}\a_{\da,-\bk} + \D_\bk^\ast \a_{\da,\bk}^\dagger\a_{\ua,-\bk}^\dagger\right),
\lb{Hsc}
\ee
with the gap function:
\be
\D_\bk = \frac 1N\sum_{\bk'}V_{\bk,\bk'}^{(\text{sc})}\langle\a_{\ua,\bk'}^\dagger\a_{\da,-\bk'}^\dagger\rangle
\lb{dk}
\ee
and the SC coupling function:
\be
\quad V_{\bk,\bk'}^{(\text{sc})} = J\g_{\bk - \bk'}\left(1 + \cos^2\l_\bk  + \cos^2\l_{\bk'} - 2\cos^2\l_\bk\cos^2\l_{\bk'}\right).
\lb{vkk}
\ee
Then, taking the normal part of this Hamiltonian from equation~\eqref{Hn} as:
\be
H_\text{n}^{(\text{sc})} = \sum_{\s,\bk} \x_\bk\a_{\s,\bk}^\dagger \a_{\s,\bk},
\lb{H0}
\ee
with the dispersion law $\x_\bk = \e_\bk - \e_{\text{F}}$, the usual diagonal form of the SC Hamiltonian:
\be
H^{(\text{sc})} = H_\text{n}^{(\text{sc})} + H_\text{an}^{(\text{sc})} = \sum_{\s,\bk} E_\bk \tilde\a_{\s,\bk}^\dagger \tilde\a_{\s,\bk},
\lb{bsc}
\ee
contains the BSC spectrum $E_\bk = \sqrt{\x_\bk^2 + |\D_\bk|^2}$ and the Bogolyubov quasiparticle operators:
\[\tilde\a_{\s,\bk} = \cos\phi_\bk\a_{\s,\bk} + \sin\phi_\bk\a_{-\s,-\bk}^\dagger,\]
with $\phi_\bk = \arctan E_\bk/|\D_\bk|$.

As known, the BCS standard routine for common metals expresses the anomalous average in the gap function, equation~\eqref{dk}, through the spectral theorem, equation~\eqref{spt}, 
where the imaginary part of anomalous GF is:
\be
\frac1\piup \Im \lGF\a_{\da,-\bk}^\dagger|\a_{\ua,\bk}^\dagger\rGF = \D_{\bk}\d\left(\e^2 - E_{\bk}^2\right)\tanh \frac{\b\e}2.
\lb{imgf}
\ee
Then, its integration in $\e$ results in the general gap equation:
\be
\D_\bk = \frac1{2N}{\sum_{\bk'}}' \frac{V_{\bk,\bk'}^{(\text{sc})}\D_{\bk'}}{E_{\bk'}}\tanh \frac{\b E_{\bk'}}2,
\lb{geq}
\ee
where the summation range in  $\sum'_{\bk'}$ is delimited by isoenergetic surfaces of equal energy distance $\o_{\text{D}}$ (the Debye energy) up and down from 
$\e_{\text{F}}$ (the BCS shell), and the SC coupling function is usually taken in the factorized form, $V_{\bk,\bk'}^{(\text{sc})} = V^{(\text{sc})} f_\bk f_{\bk'}$, with the SC coupling 
parameter $V^{(\text{sc})}$ and the SC symmetry factor $f_\bk = f(\varphi)\theta(\o_{\text{D}}^2 - \x^2)$ restricted to the BCS shell and admitting some angular dependence. 
Then, the gap function is expressed as $\D_\bk = \D f_\bk$ where the gap parameter $\D$ is found from the BCS gap equation:
\be
\frac1\l = \frac 1{2\piup} \int_0^{2\piup} \rd\varphi  f^2(\varphi)\int_0^\o \frac{\rd\x }{\sqrt{\x^2 + \D^2f^2(\varphi)}}\tanh\frac{\b\sqrt{\x^2 + \D^2f^2(\varphi)}}2
\lb{bcsg}
\ee
with the dimensionless SC coupling constant $\l = V\r_{\text{F}}$, and this should define all the SC state properties.  

However, the considered case of doped and disordered metal differs from the above classical scenario in some essential conditions: 

I) the band-like states possess a finite linewidth $\G(\e)$ whose characteristic value near the Fermi energy $\G_{\text{F}} = \G(\e_{\text{F}}) \sim \d_c$ [by equation~\eqref{irm} at 
the $\sim$ relation] can be comparable to or even greater than the expected SC spectrum gap $\D$, 

II) the inner limit for ${\sum}'$ around the M-point is defined by the mobility edge $\e_c$, since there are no mobile carriers for $\e_{\s,\bk} > \e_c$, so the BCS shell 
upper width (above $\e_{\text{F}}$), $w_u(c) = \d_{\text{F}} - \d_c$, only makes sense at $c \geqslant c_{\text{met}}$ and then it grows with $c$ as shown in figure~\ref{fig7}, 

III) the outer ${\sum}'$ limit is defined by the decay of SC coupling function and it is deduced from equation~\eqref{vkk} to be near the isoenergetic line of $\sim J$, which is 
doping independent, but the BCS shell lower width (below $\e_{\text{F}}$): $w_l(c) = J -  \d_{\text{F}}$, decreases with $c$ from a high initial value and vanishes at certain~$c_{\text{cr}}$. 

The condition I) modifies equation~\eqref{imgf} by changing there the delta-function for a Lorentzian:
\[\frac{2\e\G_{\text{F}}}{\left(\e^2 - E_\bk^2 -G_{\text{F}}^2\right)^2 + 4\e^2\G_{\text{F}}^2}\,,\]
and, after the spectral theorem integration in $\e$ (at $T = 0$), this introduces to equation~\eqref{geq} the factor:
\[1 - \frac1\piup\arctan\frac{2E_{\bk'}\G_{\text{F}}}{E_{\bk'}^2 - \G_{\text{F}}^2},\]
besides $E_{\bk'}^{-1}$, which notably differs from unity at $E_{\bk'} \sim \G_{\text{F}}$. 

The conditions II) and III) raise the question of restoring the particle-antiparticle symmetry ($\x \leftrightarrow -\x$) of the SC spectrum, equation~\eqref{Hsc}, at the original 
asymmetry of inner and outer limits in the gap equation, this asymmetry yet varying with $c$. The simplest remedy here can be sought in the choice of symmetric 
but $c$-dependent BCS shell of the energy width:
\be
w(c) = \min \left\{ w_u(c),w_l(c)\right\},
\lb{sh}
\ee
as shown in figure~\ref{fig9}.
\begin{figure}[!t]
		\center \includegraphics[width=7cm]{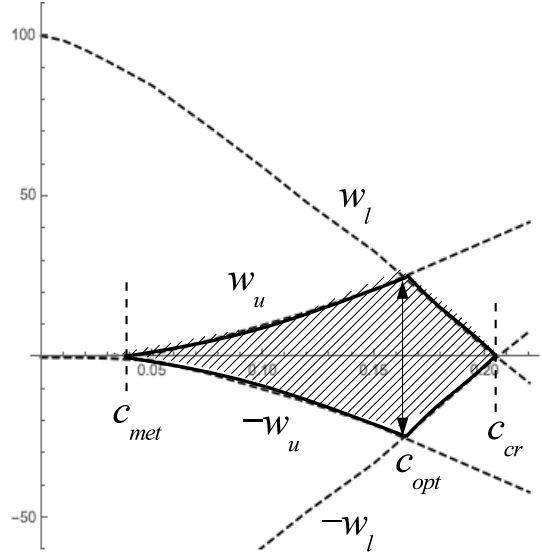}\\
	\caption{Doping dependent BCS-like energy shell around the Fermi level obtained by symmetrization of two limiting functions $w_u(c)$ and $w_l(c)$ (with their 
		mirror images) and the characteristic concentration values to produce the bell-shaped $\D(c)$ behavior.}
	\lb{fig9}
\end{figure}
\begin{figure}[!b]
	\center \includegraphics[width=7cm]{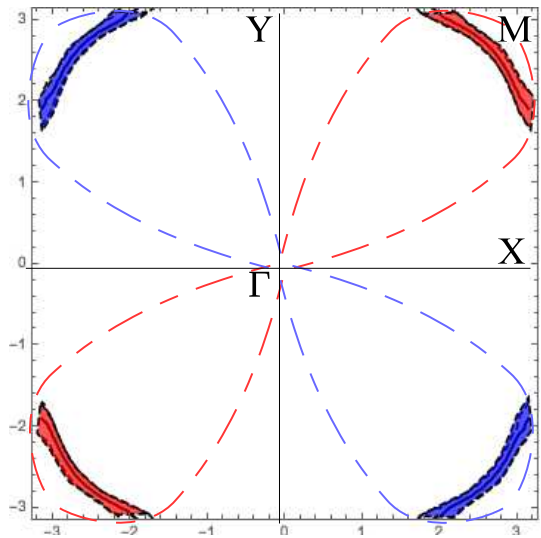}\\
	\caption{(Colour online) The $d$-wave symmetry of the SC coupling factor $f_\bk$ (and so of the gap function $\D_\bk$) when referred to the central $\G$-point in BZ. Red and blue 
		areas present respectively the ranges of its positive and negative values (in the limits defined for the doping level $c = 0.15$) and correspond to the $d$-wave 
		antinodal directions, while its zero values along $\G$-X and $\G$-Y axes do to the nodal directions.}
	\lb{fig10}
\end{figure}

Next, though there is no exact factorization in the full functional form by equation~\eqref{vkk}, it appears in the dominant part of its $\g$-term for $\bk$ and $\bk'$ near the M-point:
\[\g_{\bk - \bk'} \approx \sin\tfrac{k_x}2\sin\tfrac{k_z}2 \sin\tfrac{k'_x}2\sin\tfrac{k'_z}2,\]
while all the $\cos^2\l$ terms in this $\bk,\bk'$ range are close to unity within $O(q^4) \ll 1$ precision, defining the approximate factor function:
\be
f_\bk = \sin\tfrac{k_x}2\sin\tfrac{k_z}2,
\lb{fk}
\ee
besides the coupling parameter $V^{(\text{sc})} = J$. Then, it should be noted that the gap function $\D_\bk$ resulting from the choice of factor by equation~\eqref{fk} automatically 
presents the $d$-wave symmetry with respect to the BZ $\G$-point, since this $f_\bk$ changes its sign at different Fermi arcs (near different $C_4$ replicas of the M-point) 
as seen in figure~\ref{fig10}. Thus, this is straightly defined by the CuO$_2$ lattice geometry and by strong AFM spin correlations, getting rid of a more involved analysis on equation~\eqref{geq} with several competing candidate symmetries of the SC order. 

Further on, the factor function square, $f_\bk^2$, decays from unity in the M-point down to 1/2 at the isoenergetic line by $\approx J$ below $\e_{+,\text{max}}$, justifying its use for the 
BCS shell in the gap equation. As to the $d$-wave angular dependence for this factor, it can be neglected when integrating mainly within the central sector of one BZ quadrant 
(of one lobe in figure~\ref{fig10}).

Under these simplifications, the actual gap equation (at zero temperature) is written as:
\be
\frac 1{\l_{sc}} = \int_0^{t_c} \rd t\left(1 - \frac1\piup\arctan \frac{2\d_c\cosh t}{\cosh^2t - \d_c^2}\right),
\lb{geqq}
\ee
where $\l_{sc} = J/t'$ resulted from the expansion: $\e_{+,\bk} \approx \e_{+,\text{max}} - t'q^2$, of equation~\eqref{majmind} near the M-point and the integration upper limit is doping 
dependent: $t_c = \arcsinh [w(c)/\D]$. Its numeric solution for the gap parameter $\D$ in function of doping $c$ presents a non-monotonous dependence (figure~\ref{fig11}) 
vanishing at the terminal values $c_{\text{met}}$ and $c_{\text{cr}}$, in qualitative agreement with the experimentally observed bell-shaped $\D(c)$ curve in doped high-$T_\text{c}$ cuprate 
systems. The effect of Fermi level broadening [by the arctan term in equation~\eqref{geqq}] is most pronounced for intermediate $c$ values from the whole doping range for SC. 
\begin{figure}[!t]
	\center \includegraphics[width=10cm]{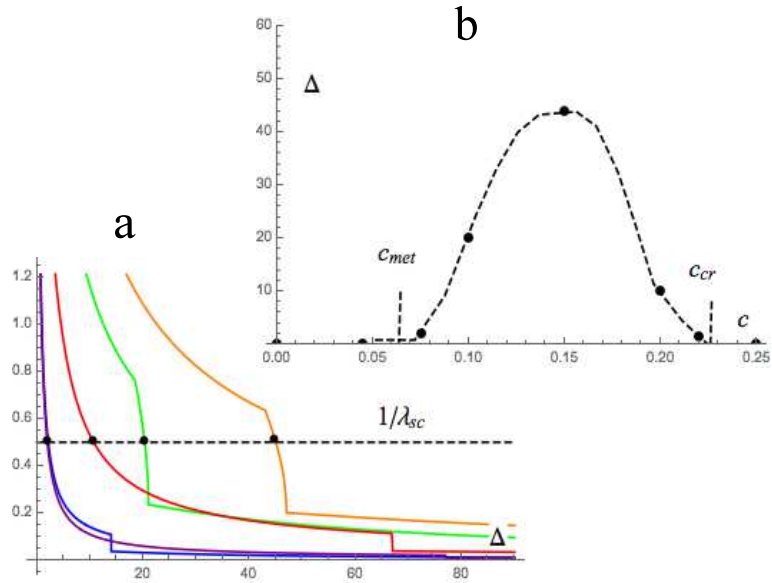}\\
	\caption{(Colour online) a) Numerical solutions of the gap equation, equation~\eqref{geqq}, for different doping levels: $c = 0.075$ (blue), $c = 0.1$ (green), $c = 0.15$ (orange), $c = 0.2$ (red), 
		$c = 0.22$ (purple). The suppression, due to the Fermi linewidth $\G_{\text{F}}$ (downsteps in the curves), is most notable at intermediate dopings (around the bell top, 
		at $c_\text{opt}$) while tending to the standard BCS solutions near the end points $c_{\text{met}}$ and $c_{\text{cr}}$. b) The resulting non-monotonous (bell-shaped) $\D(c)$ behavior. }
	\lb{fig11}
\end{figure}

Another important feature of the SC state corresponding to this solution is anomalously low superfluid density $n_s$, close to the effective weight of conductive hole states 
by equation~\eqref{wc} at not too high doping levels and vanishing at $c \to c_{\text{cr}}$, in qualitative similarity with the experimental observations in the overdoped range \cite{boz}. Moreover, 
an additional analysis shows that,  in the system normal state, there can be also conductive states in the impurity subband but they do not contribute to the SC condensate 
(this issue staying beyond the scope of the present work).

\section{Discussion and conclusions}
\lb{disc}
The above model consideration does not pretend to readily give a complete description of all the details in the multifaceted process of SC pairing in real doped cuprate 
perovskites (which are known as ``bad metals'' \cite{kei}) and its transformations with the external parameters such as temperature, magnetic field, current density, etc. Nevertheless, 
this first model approximation demonstrates that some basic features of such a process can be really determined by the specific mechanisms introduced here: 

- the spin-exchange splitting of the initial spin-independent band structure into the subbands of spin-majority and spin-minority carriers, as valence and conductance bands 
for a narrow gap insulating state of the undoped system; 

- the fundamental role of the localized impurity level produced within this bandgap by the dopant potential to give rise to a continuous impurity subband with a growing doping; 

- a strong hybridization between the spin-majority and impurity subbands that produces a significant quasiparticle weight transfer from impurity to valence subband and hence 
a transfer of the Fermi level to within the spin-majority subband, near its top;

- the essential disorder effect on the states near this Fermi level, defining the necessary condition for metallization of the initially insulating system and an anomalous closeness 
of this level to the Mott mobility edge delimiting the quasiparticle weight of conductive states;

- the strong dependence of the size and shape of the Fermi surface in the Brillouin zone on the doping level above the metallization threshold, with a tendency to be transformed 
from separate Fermi arcs to a closed Fermi line;

- the specific limitation of the ranges for SC pairing on carrier states above and below the Fermi level (an analog to the classical BCS shell), whose width is determined by the 
mobility edge in the normal spectrum combined with the common decay range of the SC coupling function;

- the $d$-wave symmetry of the resulting SC order parameter directly related to the local symmetry of two AFM sublattices, its non-monotonous dependence on the doping level, 
and an anomalously low superfluid density of the SC condensate.

Other physical properties of the proposed model, including its behavior at finite temperatures, will constitute the topic of further studies.

\section*{Acknowledgements}

V.M.L. thanks the Physics and Astronomy Department of National Academy of Sciences of Ukraine for supporting these researches within the projects No. 0117U00236 and No. 
0117U00240.

\ukrainianpart

\title{До теорії високотемпературної надпровідності допованих купратів}
\author{Ю.Г. Погорєлов\refaddr{Porto}, В.М. Локтєв\refaddr{BITP,KPI}}
\addresses{
	\addr{Porto} IFIMUP-IN, факультет фізики та 
	астрономії, Портський університет, Порто, Португалія
	\addr{BITP} Інститут теоретичної фізики ім. М.М. Боголюбова НАН України, \\ вул. Метрологічна, 14-б, 
	03143 Київ, Україна
	\addr{KPI} Національний технічний університет України «Київський політехнічний інститут імені Ігоря Сікорського», пр-т Перемоги, 37, 03056 Київ, Україна
}

\makeukrtitle

\begin{abstract}
\tolerance=3000%
Здійснено теоретичний аналіз можливих ефектів, спричинених у високотемпературних надпровідних мідних перовскітах присутністю допантів. Розраховані основні спектри спінових та електронних збуджень і спостережувані властивості купратів, насамперед поведінка параметра порядку залежно від концентрації допантів. Ключова особливість запропонованої фізичної картини полягає у формуванні всередині  діелектричної щілини вихідного недопованого матеріалу специфічної домішкової підзони, яка слугує джерелом наступних металізації і переходу у надпровідний стан з анізотропним параметром порядку.
\keywords концентрація допантів, магнітні та eлектронні спектри купратів,  надпровідність, надпровідна щілина і температура переходу
	
\end{abstract}

\end{document}